\documentclass[
	letterpaper,
	10pt,
]{CSUniSchoolLabReport}

\usepackage{graphicx}
\usepackage{pgfplots}
\usepackage{subcaption}
\usepackage{subfig}
\title{\textbf{Large-scale, Multi-pass, Two-chamber RF Atomic Magnetometer}}

\author{\textbf{D.J. \textsc{Heilman}, K.L. \textsc{Sauer}, D.W. \textsc{Prescott}, C.Z. \textsc{Motamedi}}\\\textit{George Mason University, Fairfax, VA, USA} \and \textbf{N. \textsc{Dural}, M.V. \textsc{Romalis}}\\\textit{Princeton University, Princeton, NJ, USA} \and \textbf{T.W. \textsc{Kornack}}\\\textit{Twinleaf LLC, Plainsboro, NJ, USA}}

\date{\today}

\begin{document}

\maketitle

\begin{abstract}

    We describe one of the largest radio-frequency RF atomic magnetometers presently operating. A total atomic volume of 128 \(\mathrm{cm^3}\), with correspondingly large number of \(^{87}\)Rb atoms, can reduce atom noise. A total of 44 passes of the probe beam reduces photon-shot noise. The atomic vapor is divided between two chambers allowing for pumping of the cells individually; doing so with opposite-helicity light enables use as an intrinsic gradiometer. In this configuration, common-mode noise sources including light-shift noise can be reduced. Magnetic tuning fields can also be applied to the chambers individually, allowing simultaneous measurement of two frequencies. An application of this is in the search for contraband materials using Nuclear Quadrupole Resonance (NQR), for which simultaneous measurement can significantly reduce search times. We demonstrate dual-frequency measurement on an effective range of 423-531 kHz, corresponding to the NQR frequencies of ammonium nitrate NH\(_4\)NO\(_3\) at the lowest value and potassium chlorate KClO\(_3\) at the highest. We explore fundamental, as well as instrumental, noise contributions to the sensitivity in this system.

\end{abstract}
    
\section{Introduction}

    Optically-pumped atomic magnetometers are useful tools for ultra-sensitive measurement of RF magnetic fields \cite{savukov2007,dang2010,budker2014,scholtes2016,wang2018,batie2018,bevan2018,clancy2021,savukov2022,maddox2023,motamedi2023}, capable of sub-fT/\(\sqrt{\mathrm{Hz}}\) sensitivity. This is owed in part to the preciseness of the resonance condition imposed on the atomic medium by an externally applied magnetic field. Consequently, the sensitive frequency range is very narrow, making atomic magnetometers an inefficient choice for broad-band applications. We demonstrate one solution to this, by having a probe beam pass through multiple atomic volumes, each with distinct resonant fields. The probe beam encodes each corresponding frequency simultaneously, improving effectiveness in practical situations in which multiple frequencies are to be detected.\\
    Furthermore, atomic gradiometers are useful for interference rejection \cite{cooper2022,kamada2015,perry2020,zhang2020}, and accurate because they do not inductively or capacitively couple to their environment or each other. In spite of this, typical methods of atomic gradiometer measurement face practical limitations. Post-processed gradiometers are limited in their dynamic range by the receiver \cite{cooper2022,cooper2018}; while alternatively, optical subtraction of signal by repeated passing of the probe through a half-wave plate introduces light-shift noise with each pass \cite{cooper2022}. We introduce a solution to both problems: opposite-helicity pump light. By dividing the atomic medium between two cells, optically pumped with opposite-helicity circular light, the probe beam receives opposite signals from either cell, automatically cancelling common-mode signals as they are created.\\
    Previous atomic magnetometers have made sensitive measurements with large cells and few passes of the probe beam \cite{lee2006}, dominated by photon-shot noise; as well as with smaller cells and many passes, dominated by atom noise \cite{cooper2022}. By making many passes through large cells, we combine the desirable aspects of both regimes in order to balance the two and improve overall sensitivity. An in-depth discussion of the fundamental noise contributions can be found in references \cite{cooper2022,savukov2005}. In the system described here, the fundamental noise is limited by atom noise
    \begin{equation}
        \delta B = \frac{1}{\gamma}\sqrt{\frac{8}{F_z nVT_2}},
    \end{equation}
    where \(\gamma\) is the atoms' gyromagnetic ratio, \(V\) is the volume of atoms, \(F_z\) is the expectation value of the spin along the DC tuning field, \(n\) is the atomic number density, and \(T_2\) is the transverse-decay constant. Experimental sensitivity is evaluated as
    \begin{equation}    
        \text{Sensitivity}=\frac{B_{RF}}{\text{SNR}}\sqrt{t_{acq}},
    \end{equation}
    where \(B_{RF}\) is the strength of the excitation field, \(t_{acq}\) is the duration of the data acquisition, and SNR is the signal-noise-ratio, evaluated from the Fourier-domain peak at the resonance frequency.
    
\section{Experimental}

    \subsection{Setup}

        As illustrated in Fig. 1, the magnetometer consists of two 4 cm cubic anodically bonded glass cells, each containing samples of isotopically enriched \(^{87}\)Rb along with 182 Torr of N\(_2\) as a buffer/quenching gas. The cells are electrically heated to 140\(^{\circ}\)C using nonmagnetic resistance wire lining a boron-nitride oven. The atoms are optically pumped at the D\(_1\) line through windows in the oven with a pulsed broad-beam laser, divided between the cells by a beam-splitter cube. Transverse-spin polarization is measured with a multi-pass linearly polarized probe beam via Faraday rotation. The probe beam makes 22 passes through each cell (44 in total), tracing the path shown in Fig. 1 as it alternates between the cells. It walks across the entrance mirror with each return, ultimately escaping to a balanced polarimeter \cite{chauvat1997} which feeds the signal to a phase-sensitive spectrometer. The probe entrances to the oven are left open to prevent loss of light. Anti-reflective coatings on all optical faces of the cells maximize probe light transmission to roughly 1 mW (15\% of input power). To expand the probed cell volume, curved mirrors contract and diverge the probe beam such that it is broad at the cell and point-like at the entrance mirror.
        \begin{figure}[H]
            \centering
            \includegraphics[scale=0.73]{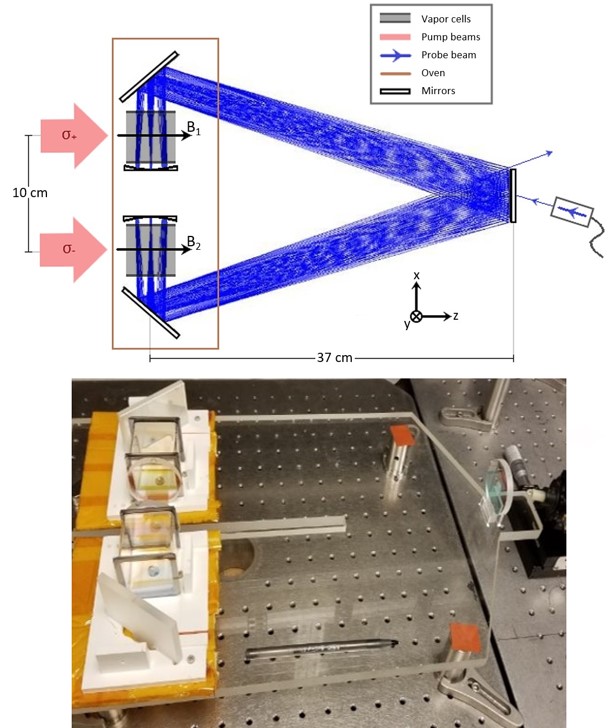}
            \caption{Optical setup of the atomic magnetometer, with schematic above and photo below. The cells are seated on a boron-nitride plate which makes up the base of the oven. The oven (not shown) has glass windows at the pump beam entrance, but is open to the probe beam. For such an open setup, convection currents must be carefully controlled so as to not deform the probe beam. }
            \label{fig:enter-label}
        \end{figure}
        
        As shown in Fig. 2, the oven is enclosed in multiple sets of highly homogeneous square magnetic field coils used to set the resonance frequencies. A set of "large-area" LA coils generates a common-DC field, while two sets of "per-sensor" PS coils generate differential-DC and rf-excitation fields. Each set consists of six coils: three homogeneous (\textit{x, y, z}) and three first-order gradients (\(\frac{dx}{dz}, \frac{dy}{dz}, \frac{dz}{dz}\)). The magnetometer is placed within a cylindrical \(\mu\)-metal shield, the inner surface of which is lined with a copper mesh to shield from Johnson noise produced in the \(\mu\)-metal. All DC tuning fields are applied in the +\textit{z}-direction, along the pump beam. Both cells' rf-excitation fields are applied in the +\textit{y}-direction.
        \begin{figure}[H]
            \centering
            \includegraphics[scale=0.4]{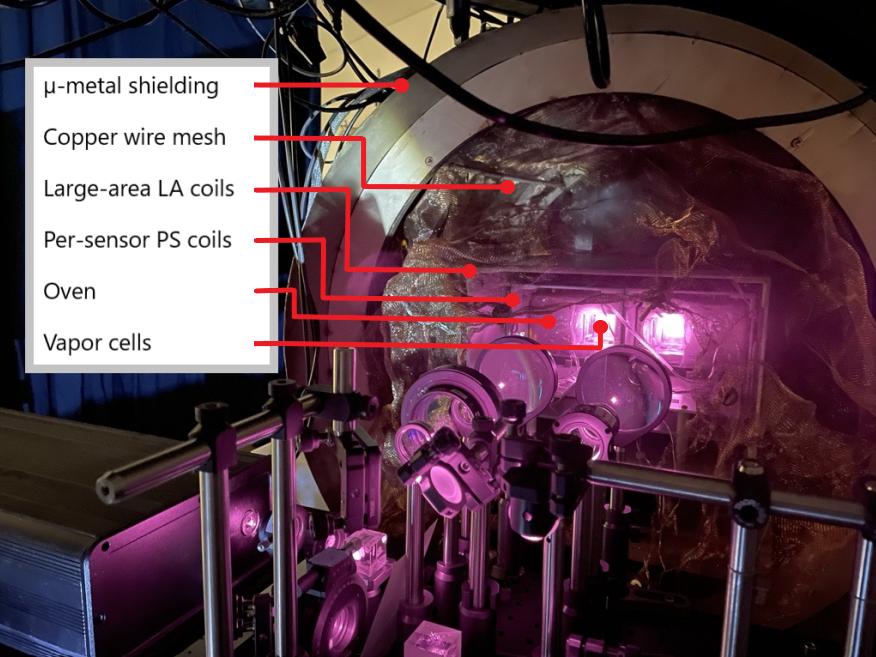}
            \caption{Experimental setup of the atomic magnetometer. The optical pumping laser and optics are shown in the foreground. Magnetic shielding with end-cup open is shown.}
            \label{fig:enter-label}
        \end{figure}

        As shown in Fig. 1, the magnetometer functions as an intrinsic gradiometer when the cells are pumped with circular light of opposite helicity. In this configuration, the Faraday rotation contributions from each cell optically subtract, cancelling common-mode noise sources.

    \subsection{Characterization}
            
        To verify the accuracy of the magnetometer, it is necessary to calibrate the magnetic fields local to the atomic vapor. To do so, we conducted an electron-spin-resonance ESR experiment. The transverse magnetization component \(M_T\) can be expressed as
        \begin{equation}
        M_T=M_0\sin(\theta_M)=M_0\sin\left(\frac{\gamma B_{RF} t_p}{2}\right),
        \end{equation}
        where \(\theta_M\) is the tipping angle of the magnetization vector, \(\gamma\) the atoms' gyromagnetic ratio, \(B_{RF}\) the magnitude of the RF pulse, and \(t_p\) the duration of the RF pulse. We confirm the strength of the applied RF field \(B_{RF}\) received by the cells by observing \(M_T\) as a function of \(t_p\), maintaining the strength of \(B_{RF}\). This is done for several RF resonant frequencies across the tested range of the magnetometer (423-531 kHz) to confirm frequency-independent signal strength. Sample data for two choices of \(B_{RF}\) frequency are given in Fig. 3.

        \begin{figure}[H]
            \centering
            \includegraphics[scale=0.5]{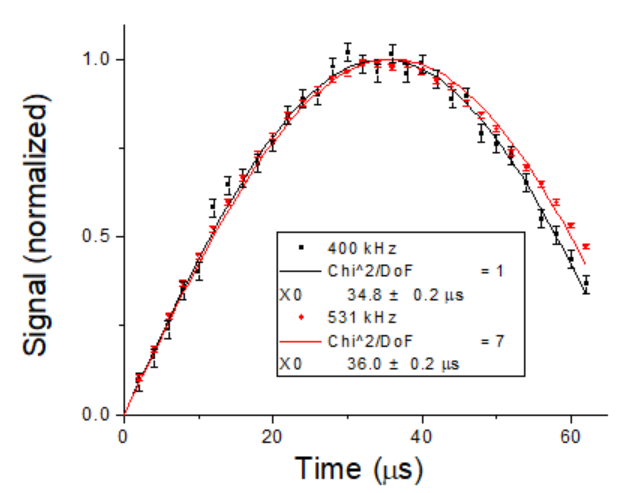}                
            \caption{\(M_T(t_p)\) for 400 and 531 kHz RF frequencies. Data suggests consistency of \(B_{RF}\) with Eq. (3) as received by the cells. Between signals, values are consistent to within 4\%. Different cells were used for either signal.}
            \label{fig:enter-label}
        \end{figure}

        In addition to the fields, we confirm the rubidium number density at operating temperature (140\(^{\circ}\)C), as well as spin-polarization rates. Estimates of these quantities in each cell are found by analyzing \(T_2\) decay for different optical pumping rates. In the limit of low polarization, atomic number density can be determined from \(T_2\) values \cite{savukov2005}. When fully pumped, and thus maximally polarized, line-narrowing of the Fourier-domain signal is used to estimate spin-polarization rates \cite{sheng2013}. Figure 4 shows typical \(T_2\) values for each cell as a function of free-induction-decay (FID) signal amplitude, which is varied with the degree of pump light allowed to reach the cells.
        \begin{figure}[H]
            \centering
            \includegraphics[scale=0.6]{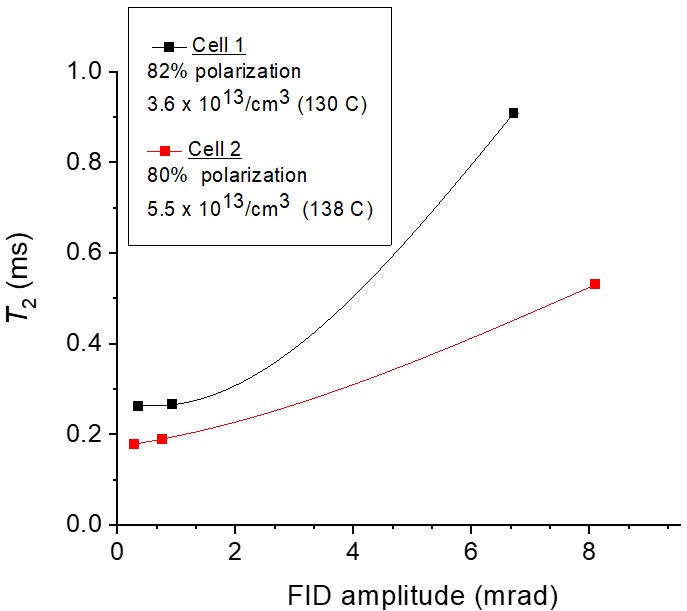}
            \caption{\(T_2\) decay constant for various optical pumping rates for each cell. Fast decay rates for unpolarized atoms reveal high atomic number densities: 3.6\(\times10^{13}\)/\(\mathrm{cm^3}\) and 5.5\(\times10^{13}\)/\(\mathrm{cm^3}\) for cells 1 and 2 respectively. When pumped, line-narrowing of the Fourier-domain signal is used to estimate spin-polarization rates \cite{sheng2013} of 80\% and 82\% for cells 1 and 2 respectively.}
            \label{fig:enter-label}
        \end{figure}

        Critical to sensitivity is the volume of the cell which is irradiated by the probe beam. This is calculated by observing the intensity of reradiated light at the D\(_2\) line. When the probe beam is sent through the unpumped atomic vapor at the D\(_1\) line, some of the re-radiated light is of the nearby D\(_2\) resonance. The D\(_2\) intensity profile, shown in Fig. 5, provides an estimate of probe width and is used to determine a total effective probing volume of 83 \(\mathrm{cm^3}\) between the cells. Re-radiation at the D\(_2\) line provides a better measure of probe interaction compared to D\(_1\), as most of the internally reflected light will be D\(_1\), obscuring the measurement. 
    
        \begin{figure}[H]
            \centering
            \includegraphics[scale=0.65]{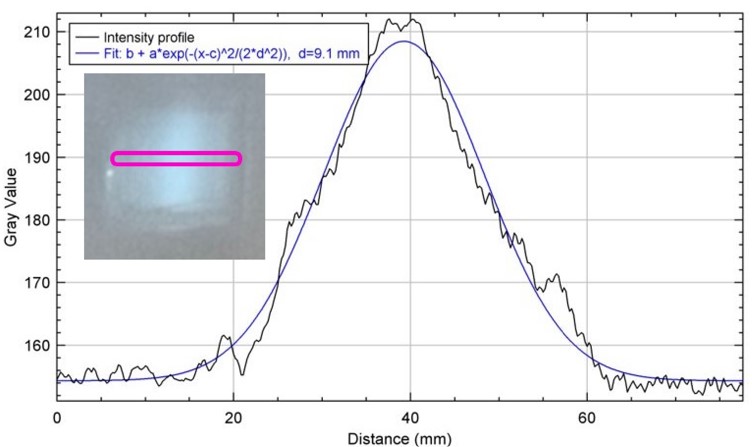}
            \caption{The effective probing volume is estimated by observing D\(_2\) fluorescence of one of the vapor cells (cross-section of the cut is shown in the inset) when probed near the D\(_1\) transition frequency. The intensity profile is fit to a Gaussian distribution and its width, as defined by the e\(^{-2}\) point, taken as that of the effective probe.}
            \label{fig:enter-label}
        \end{figure}
        
\section{Results}

    \subsection{Dual-frequency measurement}

        By applying distinct tuning and rf-excitation fields to each cell, we are able to observe two quantum resonances with a single measurement. To demonstrate this, we simultaneously measure signals from the two vapor cells as their frequencies are separated incrementally, with proper adjustment of the tuning field to maintain resonance. Both cells are initially tuned to 440 kHz via the large-area field coil. Individual adjustments are made via the per-sensor field coils in \(\pm\)20 kHz steps until their frequency difference spans the target range of the magnetometer. Two types of measurements were taken: steady-state (SS), where the test signal is turned on during data acquisition to mimic a signal of interest; and FID, where a strong, short signal is applied immediately prior to acquisition for characterization of the sensor, particularly its linewidth. These are shown in Fig. 6, along with sample time-domain data to the right illustrating the mapping of both frequencies onto a single probe beam.
        
        \begin{figure}[H]
            \centering
            \begin{subfigure}{.5\textwidth}
              \centering
              \includegraphics[scale=0.4]{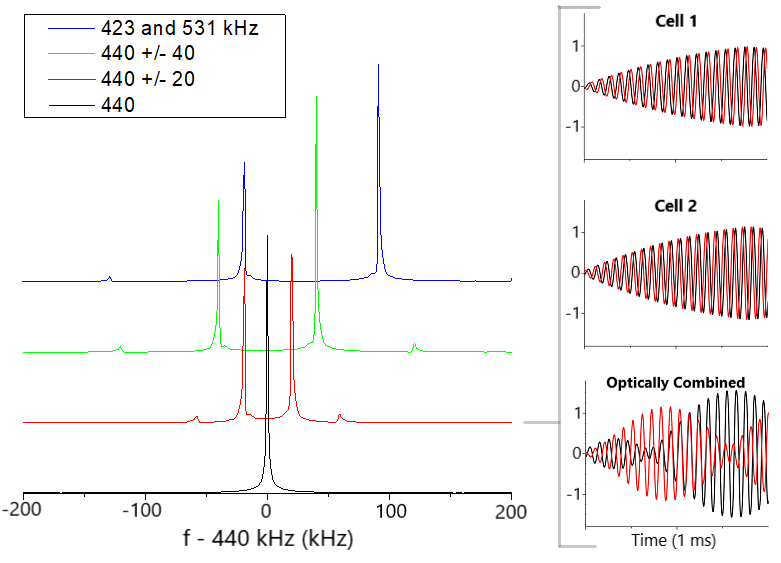}
              \caption{Steady-state SS}
              \label{fig:sub1}
            \end{subfigure}
            \hfill
            \begin{subfigure}{.5\textwidth}
              \centering
              \includegraphics[scale=0.4]{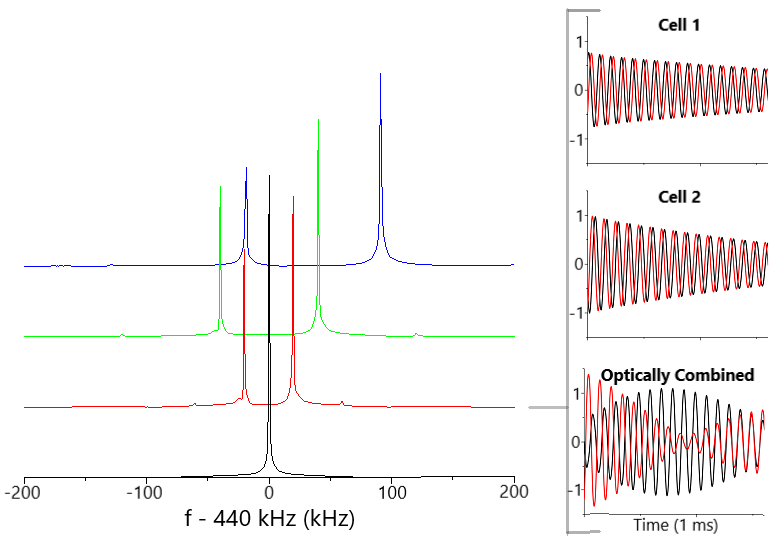}
              \caption{Free-induction-decay FID}
              \label{fig:sub2}
            \end{subfigure}
        \caption{Dual-frequency measurement for incremented frequency separation is shown in the Fourier-domain. (a) For the SS signal, the RF amplitude \(B_{RF}\)=58 pT, and RF pulse length \(t_p\)=1 ms; (b) for FID, the magnetization is initially tipped by an angle \(\theta_M\)=0.6 mrad. Time-domain data shown on the right, with the real component shown in red and the imaginary in black, corresponds to 440\(\pm20\) kHz measurement (red peaks in Fourier-domain).}
        \label{fig:test}
    \end{figure}

    Figure 6 shows some variation in signal amplitude across measurements. In principle, the magnitude of the Fourier-domain signal should be independent of frequency on the target range of 423-531 kHz. Rather than an intrinsic frequency-dependence, discrepancy between signal sizes is likely due to a misaligned pump beam, which would more strongly reduce the pumping rate for smaller tuning fields, i.e. lower frequencies. Additionally, there is a slight instability in the intensity of the pump beam, which contributes to the discrepancy. This is evidenced by the variation between the 420 (left-red) and 423 (left-blue) kHz measurements, despite similar linewidths.

    \subsection{Gradiometer through pump helicity}

        By pumping the two vapor cells with circularly-polarized light of opposite helicity, the atomic magnetometer functions as an intrinsic gradiometer, cancelling common-mode signal/noise between the cells. In this configuration, light-shift noise can, in principle, be reduced compared to existing gradiometer setups \cite{cooper2022,kamada2015,perry2020,zhang2020}. Individual field control also cancels out the nonlinear Zeeman effect, so that both cells are resonant while oppositely pumped. Overall common-mode noise is reduced 8-fold from signal addition, as shown in Fig. 7. However, noise reduction was not sufficient to have decisively reduced light-shift noise.

            \begin{figure}[H]
                \begin{tabular}{cc}
                  \includegraphics[scale=0.38]{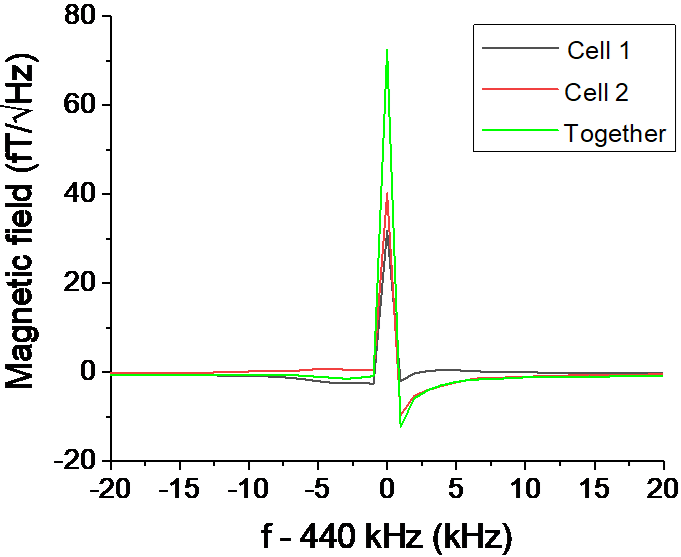} &   \includegraphics[scale=0.38]{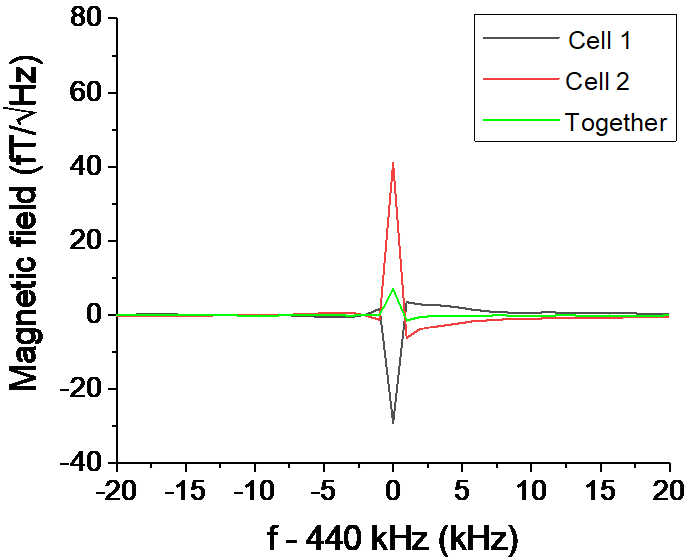} \\
                (a) Signal, +/+ pumping & (b) Signal, +/- pumping \\[6pt]
                 \includegraphics[scale=0.38]{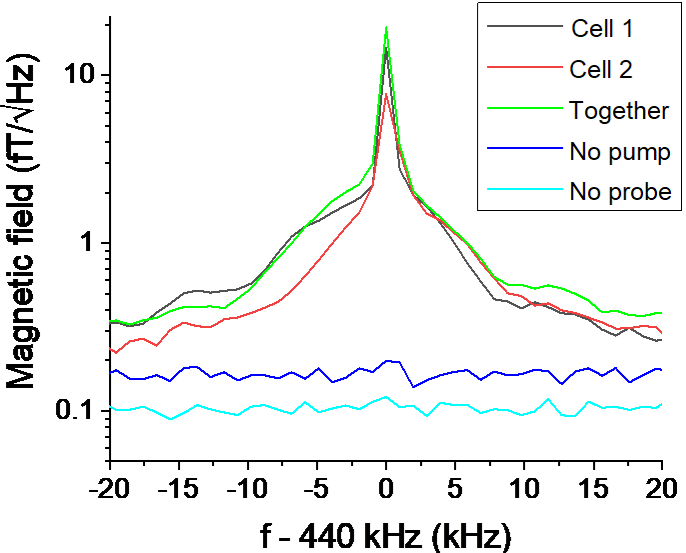} &   \includegraphics[scale=0.33]{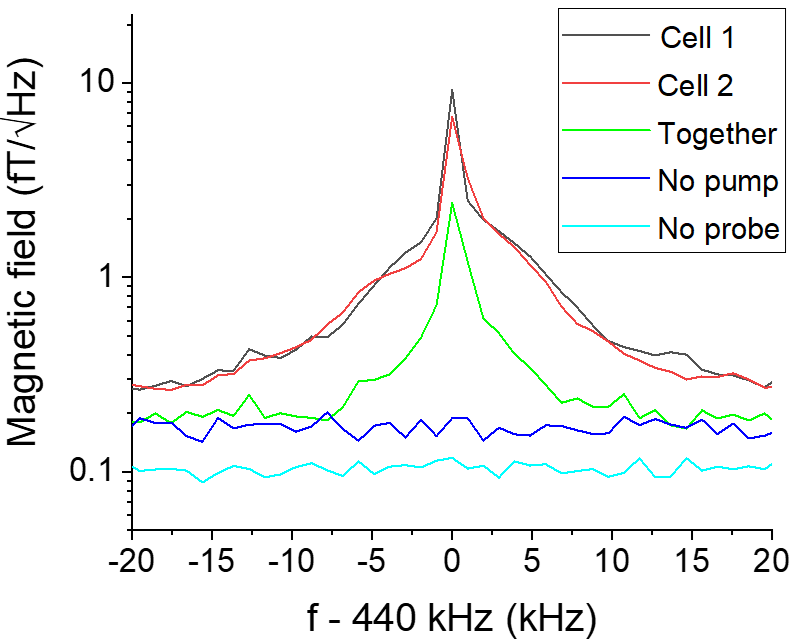} \\
                (c) Noise, +/+ pumping & (d) Noise, +/- pumping \\[6pt]
                \end{tabular}
                \caption{Magnetometer spectrum near 440 kHz, with both cells resonant at 440 kHz. An excitation of amplitude \(B_{RF}\)=1.6 pT was applied for \(t_p\)=2.048 ms. In the the upper plots, Fourier-domain signals are taken for (a) same-helicity pumping, resulting in optical addition of the cells' signals, and for (b) opposite-helicity pumping, for optical subtraction. The lower plots show the noise spectrum of the same setup, for (c) same-helicity and (d) opposite helicity pumping. Overall, we observe 10x signal reduction and 8x noise reduction for optical cancellation compared to addition.}
            \end{figure}

            Any remainder between the optically subtracted signals (b, green) is the result of differences in number density and spin-polarization between the cells. The same is true for remainder between the optically subtracted noise signals (d, green), but this discrepancy sees additional contribution from non-common-mode noise such as pump ringing, which is discussed in the following section. This is evidenced by the similar noise spectrums in each configuration for unpumped atoms. 

    \subsection{Suppression of pump ringing via phase-cycling}
    
        The limiting factor of overall sensitivity is likely ringing due to the pump laser. This describes a sudden change in the net static field direction accompanying the turning on/off of the pump beam. The circularly-polarized pump beam, which is slightly misaligned from the direction of the static field, contributes its own fictitious field arising from light-shift \cite{savukov2005}. The transient effect of it turning on/off is destabilization of the spin-polarization direction, which we observe in the time-domain signal shown in Fig. 8(a). In an ideal world the pump beam could be perfectly aligned with the static tuning field and the problem would not arise.\\
        
        To ameliorate this effect, we apply phase-cycling, wherein for successive scans, the excitation is inverted. Successive measurements then subtract , cancelling common-mode interference. Figure 8 compares time-domain noise data without (a) and with the addition of (b) phase-cycling. Phase-cycling noticeably ameliorates pump ringing, as well as general common-mode noise. Note that phase-cycling was used for previous figures.

        \begin{figure}[H]
            \centering
            \begin{subfigure}{.5\textwidth}
              \centering
              \includegraphics[scale=0.5]{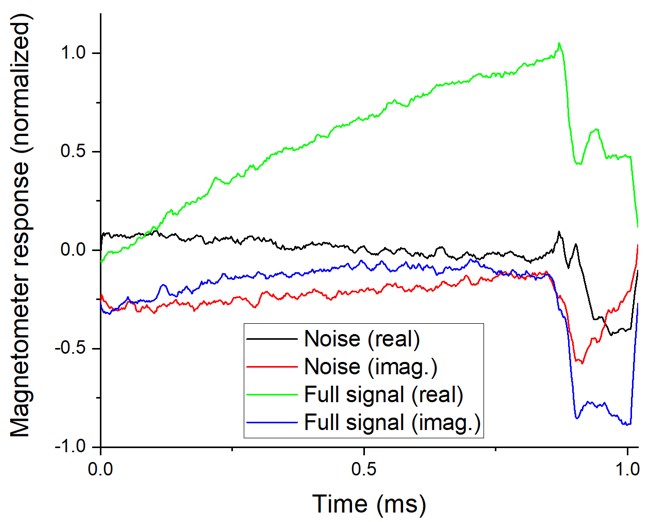}
              \caption{Without phase-cycling}
              \label{fig:sub1}
            \end{subfigure}
            \hfill
            \begin{subfigure}{.5\textwidth}
              \centering
              \includegraphics[scale=0.5]{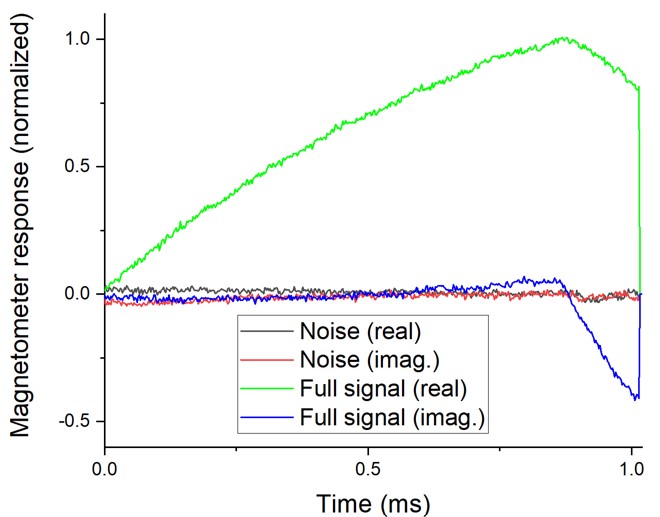}
              \caption{With phase-cycling}
              \label{fig:sub2}
            \end{subfigure}
        \caption{Ringing noise is dramatically reduced by introducing phase-cycling, as described in the text. The spike at the end of both windows is caused by transverse effects of the pump beam turning back on.}
        \label{fig:test}
    \end{figure}

    \subsection{Sensitivity}

        Without pumping of the atoms, noise sources are limited to spin-projection, photon-shot, and technical noise. To isolate the photon-shot contribution, off-resonant noise is observed with/without the probe beam. The frequency dependence of spin-projection noise distinguishes it from other noise sources. From the in-the-dark (low polarization) noise peak, the noise peak with the pump beam on, as is necessary for operation of the magnetometer, can be calculated. Using the observed linewidth, and expecting that noise power is \(\frac{2}{3}\) that of unpolarized atoms \cite{cooper2022}, we obtain a projected sensitivity of 50 aT/\(\sqrt{\text{Hz}}\). This estimate shows the potential sensitivity of the configuration if pump ringing is eliminated. A pump-laser power of 30 W maximizes sensitivity. Analysis is shown in Fig. 9 for two choices of probe-beam wavelength.

        \begin{figure}[H]
            \centering
            \includegraphics[scale=0.6]{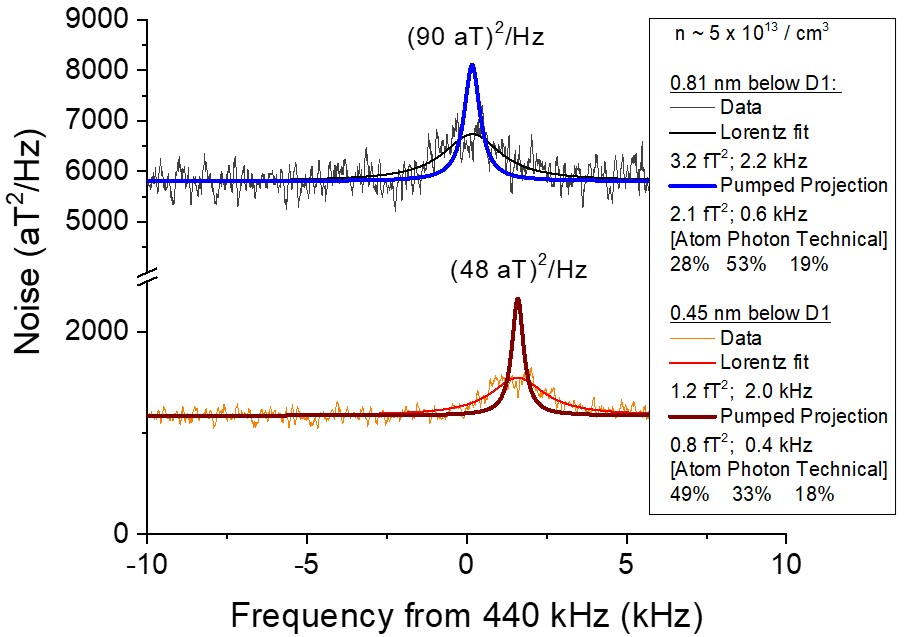}
            \caption{In-the-dark noise distributions are fit to a Lorentzian lineshape for two different probing wavelengths: 0.81 nm (blue), and 0.45 nm (red) below the D1 transition. For each, a projection of the fully-pumped noise distribution is given along with the corresponding sensitivity. This projection is much smaller than ringing noise from operation of the pump beam. Sufficient detuning of the probe beam from resonance is critical for optimal sensitivity \cite{cooper2022}.}
            \label{fig:enter-label}
        \end{figure}

        Using our measurement of the probed volume based on D\(_2\) fluorescence, and measurements of number density and polarization based on \(T_2\) decay, Eq. (1) gives a theoretical atom noise of 33 aT/\(\sqrt{\mathrm{Hz}}\). We compare this to the result shown in Fig. 9 (lower) by noting the noise contribution proportions, which reveal an experimental atom noise of 34\(\pm{3}\) aT/\(\sqrt{\mathrm{Hz}}\) and photon-shot noise of 28\(\pm{3}\) aT/\(\sqrt{\text{Hz}}\). Previous research with large cells of similar volume, but only a single pass, give comparable atom noise, but significantly larger, by a factor of four, photon-shot noise \cite{alem2013}.

\section{Conclusions}

    By varying the power of the pump and probe beams, the individual noise contributions to sensitivity can be found. We have demonstrated that by dividing the alkali-vapor between two cells with distinct magnetic fields, it is possible to simultaneously map two resonant frequencies onto a single probe beam. In principle, the number of distinct frequencies encoded on to a single probe beam can be increased by adding more cells. Such simultaneous detection would greatly speed up search speeds, for example, in the detection of contraband materials by NQR.\\
    Furthermore, we have demonstrated the ability to use opposite-helicity pump light to form an intrinsic RF gradiometer when observing a single frequency. Such optical subtraction, as opposed to post-processing subtraction from two magnetometers, can ameliorate dynamic range issues when using the gradiometer to suppress large common-mode interference. While not demonstrated here, due to experimental limitations with pump-beam pulsing, such optical subtraction also offers the possibility to reduce light shift noise.\\
    Atom noise was clearly observed in the effective atomic volume of 83 \(\mathrm{cm^3}\), although it did not dominate the noise spectra due to ringing associated with the pump beam. Atom noise of 34\(\pm{3}\) aT/\(\sqrt{\mathrm{Hz}}\) was in agreement with theoretical predictions, and was larger than photon shot noise, 28\(\pm{3}\) aT/\(\sqrt{\text{Hz}}\). This low fundamental noise demonstrates the potential of using large cells with many passes of the probe beam to make sensitive measurements.

\section{Acknowledgements}
This work was supported, in part, by the National Science Foundation (award 171118).

\printbibliography

\end{document}